\documentclass[12pt]{iopart}
\usepackage{graphicx}
\usepackage{dcolumn}% Align table columns on decimal point
\usepackage{bm}% bold math
\usepackage{tabularx}
\usepackage{ulem}
\newcolumntype{M}{>{\centering\arraybackslash}m{1.85cm}}
\usepackage[export]{adjustbox}
\usepackage{float}
\usepackage{braket}
\usepackage{makecell}
\usepackage{lipsum}
\usepackage{longtable}
\graphicspath{{figure/}}
\usepackage{xcolor}   %May be necessary if you want to color links

\usepackage{hyperref}

\begin{document}

% \begin{document}

\title[]{Investigation of entanglement in $N = Z$ nuclei within no-core
shell model}

\author{Chandan Sarma, Praveen C. Srivastava}
\address{Department of Physics, Indian Institute of Technology Roorkee, Roorkee 247667, India}	
\ead{c\_sarma@ph.iitr.ac.in, praveen.srivastava@ph.iitr.ac.in}
\vspace{10pt}

\begin{abstract}
In this work, we explore the entanglement structure of two $N = Z$ nuclei, $^{20}$Ne and $^{22}$Na using single-orbital entanglement entropy within the No-Core Shell Model (NCSM) framework for two realistic interactions, INOY and N$^3$LO. We begin with the determination of the optimal frequencies based on the variation of ground-state (g.s.) binding energy with NCSM parameters, $N_{max}$ and $\hbar \Omega$, followed by an analysis of the total single-orbital entanglement entropy, $S_{tot}$, for the g.s. of $^{20}$Ne and $^{22}$Na. Our results show that $S_{tot}$ increases with $N_{max}$ and decreases with $\hbar \Omega$ after reaching a maximum. We use $S_{tot}$ to guide the selection of an additional set of optimal frequencies that can enhance electromagnetic transition strengths. We also calculate the low-energy spectra and $S_{tot}$ for four low-lying states of $^{20}$Ne and six low-lying states of $^{22}$Na.  Finally, we calculate a few $E2$ and one $M1$ transition strengths, finding that N$^3$LO provides better results for $B(E2; 5^+_1 \to 3^+_1$) and INOY performs well for the $B(M1; 0_1^+ \to 1_1^+)$ transition in the $^{22}$Na nucleus while considering the first set of optimal frequencies.  We also observe that the second set of optimal frequencies enhances electromagnetic transition strengths, particularly for the states with large and comparable $S_{tot}$. Also, for both nuclei, the $S_{tot}$ for INOY and N$^3$LO are close while considering the second set of optimal frequencies, suggesting that the calculated $S_{tot}$ are more dependent on $\hbar \Omega$ than the interactions employed for the same model space defined by the $N_{max}$ parameter.
\end{abstract}

\section{Introduction:}
In the last two decades, the \textit{ab initio} methods have provided important benchmarks for solving nuclear many-body systems. Among several \textit{ab initio} methods, the no-core shell model (NCSM) %\cite{NCSM_r2, NCSM_p5, NCSM_p6, NCSM_p8} 
\cite{NCSM_r2,petr_PS,NCSM_p8} is very successful in describing the properties of nuclei in the low-mass region, considering all nucleons to be active and interacting via realistic $NN$ and $3N$ interactions. Like any other \textit{ab initio} methods, the NCSM also uses realistic interactions that are either derived from the meson exchange theory \cite{CDB2K3} or Quantum Chromodynamics (QCD) \cite{QCD} based on chiral effective field theory ($\chi$EFT) \cite{EFT1, EFT2, EFT3, EFT4} as an input to the many-body problem. Recently, we have explored several nuclei in the $p$- and lower $sd$-shell region within the NCSM formalism \cite{arch1, arch2, Choudhary1, Choudhary2, Choudhary3, chandan_Ne, chandan_Na}. While the calculated spectra and other nuclear observables are in good agreement with the experimental data in the case of $p$-shell nuclei, some observables are away from the actual results for the case of $sd$-shell nuclei \cite{chandan_Ne, chandan_Na}. There could be two reasons for such deficiencies: (i) the model space used for such calculations is not large enough, leaving out some important configurations, or (ii) the interactions used for such calculations lack some components that are relevant for lower $sd$-shell nuclei. Although the first one is the dominant one for our case, some minor contributions may come from the second one as well. In this work, we are focussing on the assessment of nuclear wavefunctions obtained for two different interactions corresponding to different model spaces based on quantum information techniques.  

Recently, entanglement studies for nuclear systems gained lots of importance as they can provide a new way of looking at the nuclear many-body systems \cite{QI1, QI2, QI3, QI4, QI5, QI6, QI7, QI8, QI9, QI10, QI11, QI12, QI13, QI14}. Some of these works focus on the investigation of entanglement within the nuclear shell model. In Ref. \cite{QI3}, mode entanglement and correlations within a two-nucleon system, $^{18}$O is investigated. The single orbital entanglement and mutual information are explored for $p$-shell nucleus, $^8$Be in Ref. \cite{QI4}. Kruppa \textit{et. al. } in Ref. \cite{QI5} associated mode entanglement with seniority states of atomic nuclei across different mass regions. On the other hand, considering atomic nuclei as a bipartite system of protons and neutrons, proton-neutron entanglement was investigated in Ref. \cite{QI9} for a few nuclei in the $sd$- and $fp$-region within the shell model formalism. In Ref. \cite{QI10}, a few nuclei from $p$- and $sd$-shell were explored for different entanglement measures such as single-orbital entanglement, von Neumann entropy, and mutual information. There were also a few works that combined the \textit{ab initio} methods for solving nuclear many-body problems with the quantum information techniques \cite{QI2, QI10, QI12}. On the other hand, lots of efforts are also put into the application of quantum computation for nuclear systems \cite{QI4, QC1, QC2, QC3, QC4, QC5, QC6, QC7, QC8, QC9}. In the long-term \textit{ab initio} methods, in combination with quantum computation and quantum information techniques, have the power to revolutionize the study of atomic nuclei. 

 In Ref. \cite{QI2}, the entanglement between the single-particle states in $^4$He and $^6$He g.s. was explored within the NCSM formalism using four different kinds of single-particle basis. Chiral $NN$ interaction was employed for this work. On the other hand, in Ref. \cite{QI13}, it was established that the proton-neutron entanglement within the low-energy spectra of $N = Z$ systems is higher than the $N \neq Z$ systems while using the phenomenological USDB interaction. However, no such increment was observed while using random two-body interactions \cite{QI13}, suggesting its dependence on the interactions used. In this work, we are interested to test the dependence of single orbital entanglement entropies on the interactions employed within the NCSM formalism, taking the example of two $N = Z$ nuclei: $^{20}$Ne and $^{22}$Na. Additionally, we aim to test the dependence of this entanglement measure on the NCSM parameters: $\hbar \Omega$ and $N_{max}$ and possible links with the electromagnetic transition strengths. 

This paper is organized as follows: in section \ref{sect2}, some basics of the NCSM formalism, along with a brief discussion on the $NN$ interactions employed in this work, are given. The single orbital entanglement entropy used in this work is also briefly described in this section. The NCSM results for energy spectra, single orbital entanglement, and electromagnetic transition strengths are discussed in section \ref{sect3}. Finally, we summarize our work in section \ref{sect4}.

\section{Formalism:}
\label{sect2}
The NCSM is a leading \textit{ab initio} method that treats all nucleons in a nucleus as active point particles interacting through realistic $NN$ and $NN + 3N$ interactions. In this study, we use only the $NN$ interactions, with the Hamiltonian expressed as follows \cite{NCSM_r2}:
% The NCSM is one of the prominent \textit{ab initio} methods in which all nucleons of a nucleus are considered to be active point particles interacting via realistic $NN$ and $NN$ + $3N$ interactions. In this work, we are using only the $NN$ interactions for which the Hamiltonian can be written as follows \cite{NCSM_r2}:
\begin{equation}
%\begin{align}
\label{eq:(1)}
H_{A}= T_{rel} + V = \frac{1}{A} \sum_{i< j}^{A} \frac{({\vec p_i - \vec p_j})^2}{2m}
+  \sum_{i<j}^A V^{NN}_{ij}, %+ \sum_{i<j<k}^{A} V_{NNN,ijk} 
%+\sum_{i<j<k}^A V^{NNN}_{ijk}
% \end{align}
\end{equation}
where, $T_{rel}$ is the relative kinetic energy and $V^{NN}_{ij}$ is the $NN$ interaction containing nuclear part as well as Coulomb part.

In the NCSM approach, the non-relativistic Schrödinger equation for an A-nucleon system is solved through large-scale matrix diagonalization in a many-body Harmonic oscillator basis. This basis is built using Slater determinants (\textit{M}-scheme) of single-particle Harmonic oscillator orbitals with a truncation parameter, $N_{max}$, which shows the number of major shells taken above the minimum configuration allowed by the Pauli exclusion principle. The \textit{M}-scheme basis with $N_{max}$-truncation enables a smooth separation of the center-of-mass and the relative coordinates. In order to achieve converged results with hard-core potentials that have strong short-range correlations, large $N_{max}$ calculations are necessary, but these calculations are computationally demanding. Renormalization techniques, such as the Okubu-Lee-Suzuki (OLS) \cite{OLS1, OLS2, OLS3, OLS4} transformation and the similarity renormalization group (SRG) \cite{SRG1, SRG2}, are used to soften these potentials, allowing for converged results in computationally feasible model space. In this study, the OLS technique is used to derive an effective Hamiltonian.

The Hamiltonian shown in Eq. \ref{eq:(1)} is in relative co-ordinate and to use it in the Harmonic oscillator basis and to facilitate the derivation of the OLS effective Hamiltonian, the center-of-mass (c.m.) Hamiltonian, $H_{c.m.}$ is added to the original Hamiltonian in Eq. \ref{eq:(1)}: 
	
\begin{equation}
	\label{eq:(2)}
\hspace{-2cm}	H^{\Omega}_A = H_A + H_{CM} = \sum_{i = 1}^A [\frac{p_i^2}{2m} + \frac{1}{2}m \Omega^2 r_i^2]
	+ \sum_{i < j}^A [V_{NN, ij} - \frac{m \Omega^2}{2 A} (r_i - r_j)^2].
	%\quad
\end{equation}
Here, $H_{c.m.}$ = $T_{c.m.}$ + $U_{c.m.}$ and $T_{c.m.}$ and $U_{c.m.}$ are the kinetic and potential terms for the center of mass coordinate. $H_A$, being translationally invariant which does not change the intrinsic properties once $H_{c.m.}$ is added to it.

To develop effective interactions, the infinite HO basis is first divided into $P$- and $Q$-spaces using projection operators, where $P$-space contains all the HO orbitals allowed by the truncation parameter $N_{max}$, and $Q$-space consists of the excluded orbitals. The effective Hamiltonian is then constructed via the OLS unitary transformation applied to $H^{\Omega}_A$ (as shown in Eq. \ref{eq:(2)}). Although the OLS transformation starts with the $NN$ Hamiltonian (Eq. \ref{eq:(1)}), it introduces up to \textit{A}-body terms. In this study, we focus only on 2-body terms, as they contribute more significantly than higher-order many-body terms. Finally, the $H_{c.m.}$ is subtracted, and the Lawson projection term, $\beta (H_{\mbox{c.m.}} - \frac{3}{2}\hbar\Omega)$ \cite{Lawson}, is added to the Hamiltonian. Here, $\beta$ is set to 10, and the inclusion of the Lawson projection pushes the energy levels that arise from CM excitations upward. Thus, the final form of the Hamiltonian is:

% \begin{equation*}
% \label{eq:(3)}
% \hspace{-2cm}{\small H_{A,\rm eff}^{\Omega} = P\left\{ \sum _{i<j}^{A} \left[ \frac{{(\vec p_i - \vec p_j)}^2}{2mA} + \frac{m {\Omega}^2}{2A} {(\vec r_i - \vec r_j)}^2 \right] +\sum_{i<j}^{A} \left[ V^{NN}_{ij} - \frac{m {\Omega}^2}{2A}{(\vec r_i - \vec r_j)}^2\right]_{\rm eff} \\ 
% \quad
% \end{equation*}

% \begin{equation}
% \label{eq:(3)}
% {\small
% + \beta \Bigg(H_{\mbox{c.m.}} - \frac{3}{2}\hbar\Omega \bigg) \Bigg\} P \ 
% \quad
% \end{equation}

\begin{equation*}
\label{eq:(3)}
\hspace{-2cm}{\small H_{A,\rm eff}^{\Omega}} = P \Bigg \{ \sum _{i<j}^{A} \Bigg [ \frac{{(\vec p_i - \vec p_j)}^2}{2mA} + \frac{m {\Omega}^2}{2A} {(\vec r_i - \vec r_j)}^2 \Bigg ] + \sum_{i<j}^{A} \left[ V^{NN}_{ij} - \frac{m {\Omega}^2}{2A}{(\vec r_i - \vec r_j)}^2\right]_{\rm eff}
\end{equation*}

\begin{equation}
\label{eq:(3)}
{\small
+ \beta \Bigg (H_{\mbox{c.m.}} - \frac{3}{2}\hbar\Omega \Bigg) \Bigg\} P }
\end{equation}

The effective Hamiltonian in Eq. \ref{eq:(3)} depends on the number of nucleons (A), the HO frequency ($\Omega$), and the number of HO orbitals included in the $P$-space, which is determined by the truncation parameter $N_{max}$. To reduce the influence of many-body terms when constructing an effective Hamiltonian, a large model space is required for NCSM. In the extreme limit where $N_{max} \rightarrow \infty$, the effective Hamiltonian from Eq. \ref{eq:(3)} approaches the bare Hamiltonian of Eq. \ref{eq:(1)}, and the NCSM results with the effective Hamiltonian converge to the exact solution.

In this work, we employed two realistic interactions, namely, inside non-local and outside Yukawa (INOY) \cite{INOY} and next-to next-to-next-to-leading order (N$^3$LO) \cite{N3LO}. The INOY interaction has a short-range non-local part for r $\le$ 1 \textit{fm} and at large distance $\ge$ 3 \textit{fm}, it is the same as the Yukawa tail of the Argonne v18 (AV18) interaction \cite{av18}. The short-range part is then evaluated by fitting $NN$ data and the binding energy of $^{3}$He. On the other hand, the N$^3$LO interaction is obtained from the fourth order of chiral perturbation theory. The long-range part of the N$^3$LO interaction is associated with the pion exchange, while the short-range interactions are defined in terms of contract terms. There are 29 parameters in N$^3$LO interaction, and those are fitted to reproduce $NN$ data up to 290 MeV. These two interactions were employed within the NCSM formalism discussed above to study $Z = N$ nuclei, $^{20}$Ne and $^{22}$Na. 

The atomic nuclei being one of the prominent many-body systems, entanglement studies can provide lots of information about the underlying structure. Quantum entanglement defines relationships between distinct segments within a system that cannot be independently described. When two systems, denoted as $A$ and $B$, each with states represented by $|\psi_A \rangle$ and $|\psi_B \rangle$ respectively, are entangled, their combined state $|\psi \rangle$ cannot be expressed as the simple tensor product of their individual states. Among different entanglement metrics, we are interested in exploring the single orbital entanglement within the NCSM formalism due to its direct relationship with the single orbital occupancies. When computing the entanglement of a single orbital, a division of the orbital from the remaining of the system is considered. This is expressed as:

\begin{equation}
S_i = -\gamma_i \times \log_2 \gamma_i - (1- \gamma_i) \times \log_2 (1- \gamma_i),    
\end{equation}
where $\gamma_i = \langle \psi | a_i^\dagger a_i | \psi \rangle$ denotes the occupation number of the $i^{th}$ orbital in the state $|\psi \rangle$ \cite{QI11}. Another pertinent quantity in this context is the total single-orbital entanglement, denoted as 
\begin{equation}
    S_{tot} = \sum_i S_i = - \sum_i  [\gamma_i \times \log_2 \gamma_i + (1- \gamma_i) \times \log_2 (1- \gamma_i)].
\end{equation}

\section{Results and Discussions:}
\label{sect3}
The NCSM calculations for medium mass nuclei become computationally difficult mainly due to the huge dimensions of the Hamiltonian matrix. In the case of $^{20}$Ne and $^{22}$Na, the g.s. are 0$^+$ and 3$^+$, respectively and 7.5 $\times$ 10$^7$ and  5.2 $\times$ 10$^8$ are their respective \textit{M}-scheme dimensions  while considering $N_{max}$ = 4 model space. On the other hand, these dimensions become 4.4 $\times$ 10$^{9}$ and 3.0 $\times$ 10$^{10}$ for $N_{max}$ = 6, which are away from our current computational limit. Hence, all the calculations presented in this work are up to $N_{max}$ = 4, comprising 28 HO orbitals.

\subsection{Variation of ground state energy and single orbital entanglement with NCSM parameters:}

\begin{figure}
\centering
\includegraphics[scale = 0.7]{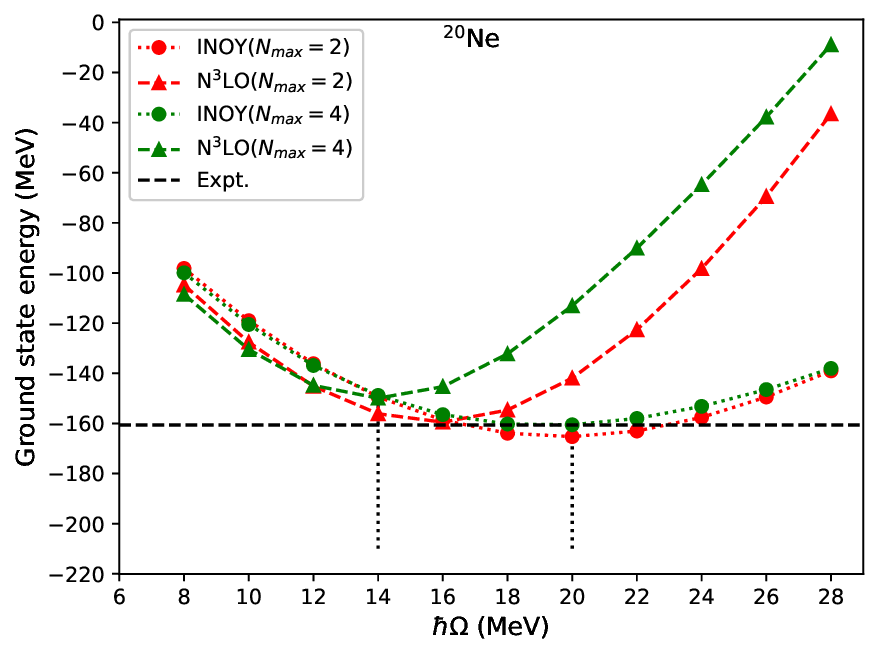}
\includegraphics[scale = 0.7]{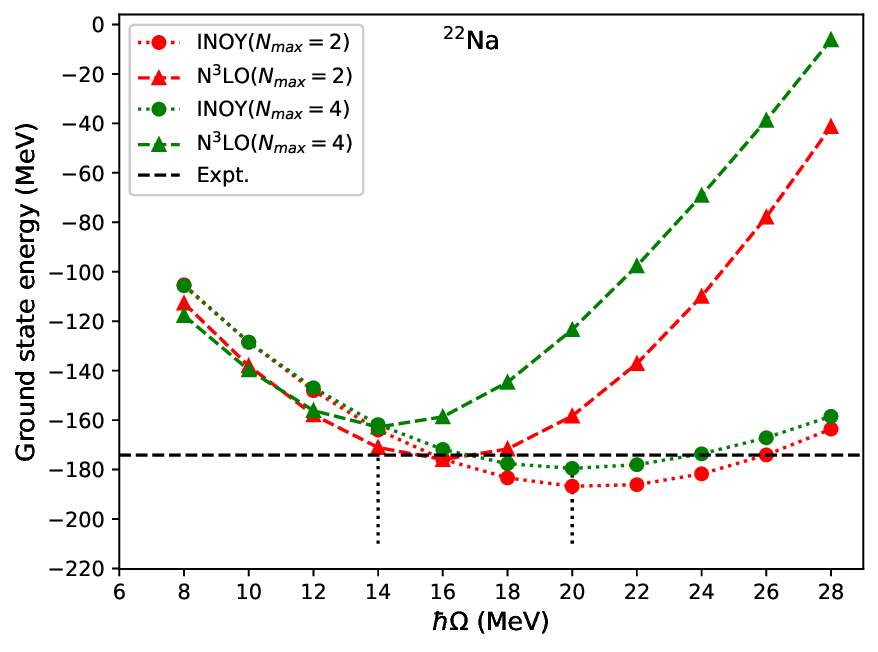}
\caption{The variation of calculated g.s. energies of $^{20}$Ne and $^{22}$Na with HO frequencies for different $N_{max}$ are shown corresponding to two different $NN$ interactions. The horizontal lines show the experimental binding energies of the g. s., and the vertical dotted lines show the optimum frequencies.}
\label{gs}
\end{figure}

The first step of the NCSM calculation is to decide an optimum frequency for which different observables are calculated later on. 
The optimum frequency for a particular interaction is decided by plotting the ground state energies for a particular interaction corresponding to different $N_{max}$. Subsequently, the frequency corresponding to the minimum g.s. energy calculated at the highest $N_{max}$ is taken as the optimum frequency for a particular interaction. For this work, the NCSM calculations for the g.s. of $^{20}$Ne and $^{22}$Na are performed for three different model spaces, $N_{max}$ = 0, 2, 4 for $\hbar \Omega$ varying from 8 to 28 MeV in steps of 2 MeV. In the first panel of Fig. \ref{gs}, the g.s. energies of $^{20}$Ne isotope are shown with realistic $NN$ interactions: INOY, and N$^3$LO for $N_{max}$ = 2 and 4 model spaces. Similarly, the same is shown in the second panel of Fig. \ref{gs} corresponding to $^{22}$Na. The optimum frequencies for INOY and N$^3$LO are 20 and 14 MeV, respectively, as shown by the vertical dashed lines in the figure. From Fig. \ref{gs}, it is evident that the g.s. binding energies do not exhibit variational behavior concerning the truncation parameter, $N_{max}$, and the harmonic oscillator frequency, $\hbar \Omega$. While the NCSM results with bare interactions are variational, converging from above with $N_{max}$ considering $\hbar \Omega$ as the variational parameter \cite{NCSM_r2}, the results for OLS transformed interactions are not variational as some part of the bare interactions is omitted.

% \begin{figure}[t]
% \centering
% \includegraphics[scale = 0.8]{22Na_gs.eps}
% \caption{The variation of calculated g.s. energies of $^{22}$Na with HO frequencies for different $N_{max}$ are shown corresponding to two different $NN$ interactions. The horizontal lines show the experimental binding energies of the g. s., and the vertical dashed lines show the optimum frequencies.}
% \label{gs}
% \end{figure}

\begin{figure}
% \centering
\hspace{-1.85cm} \includegraphics[scale = 0.63]{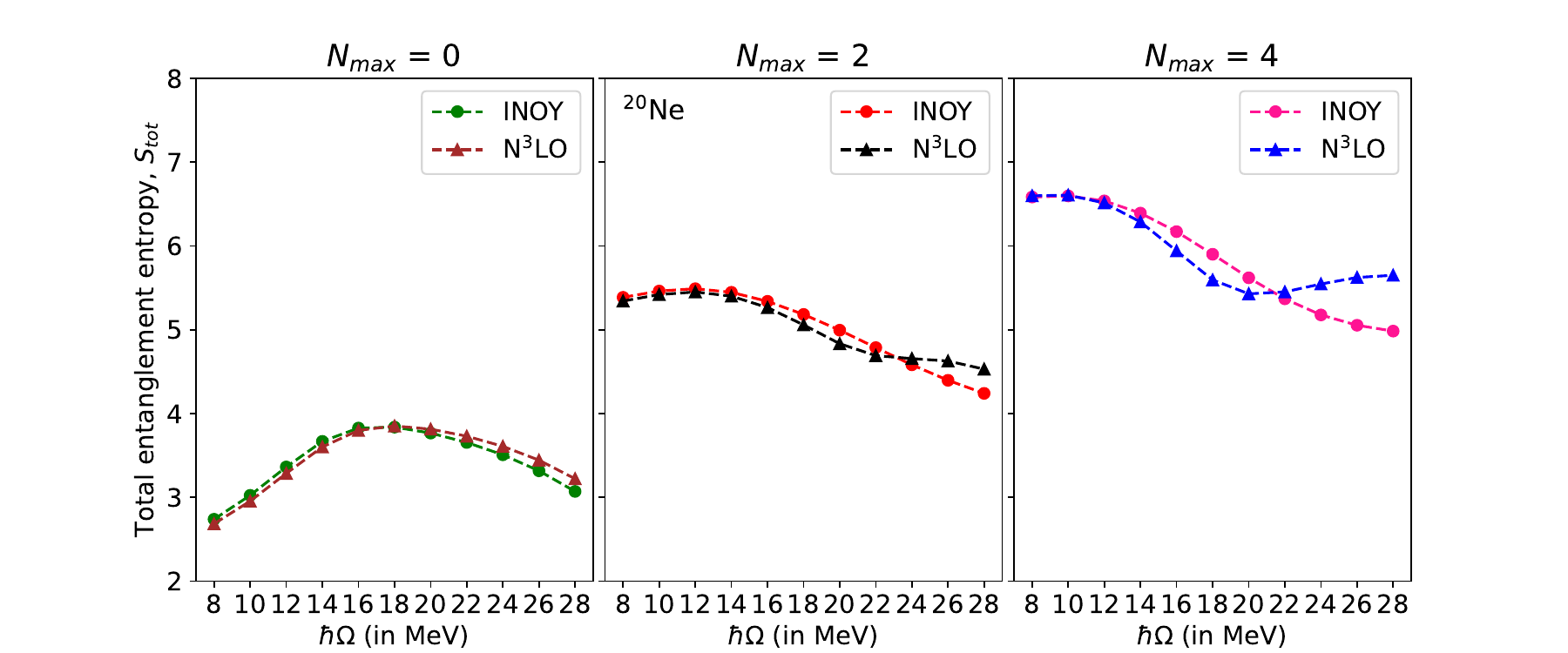}

\hspace{-1.85cm}    \includegraphics[scale = 0.63]{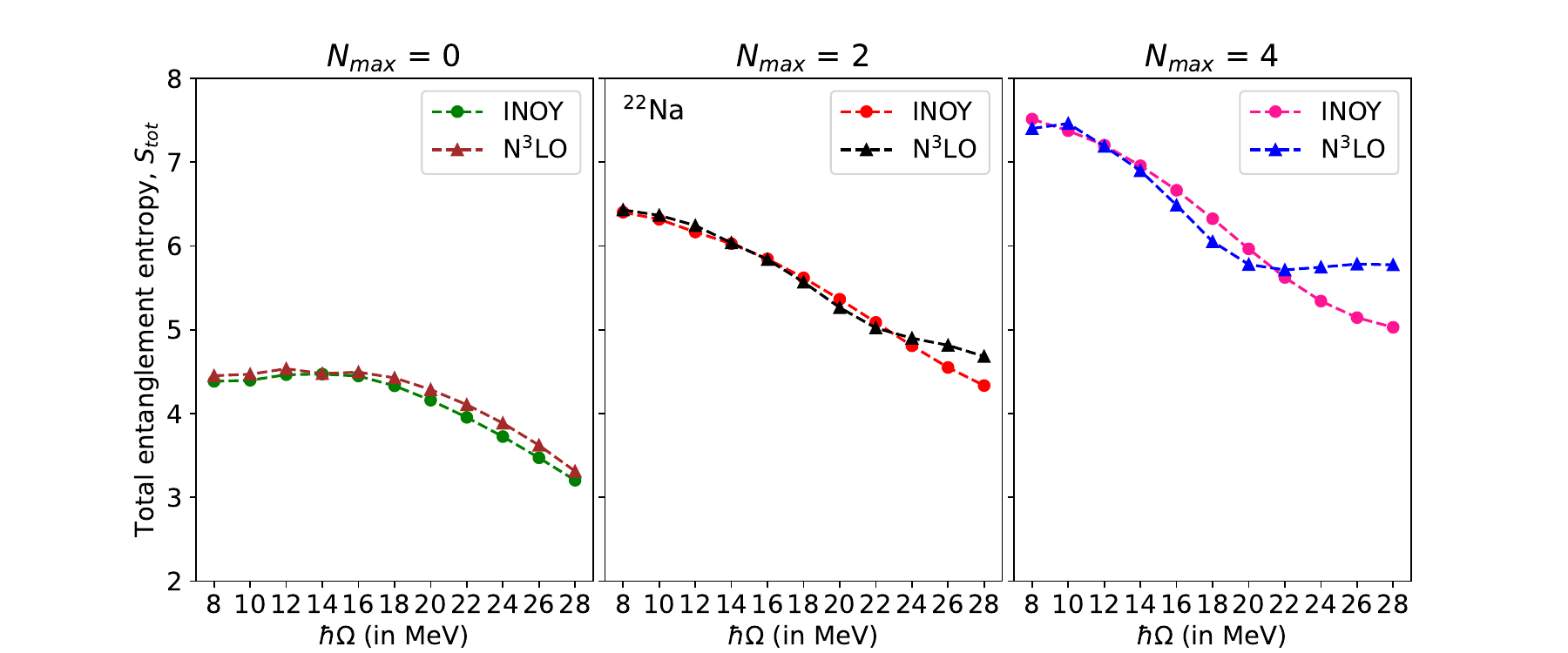}\\

\caption{The variation of calculated g. s. $S_{tot}$ of $^{20}$Ne and $^{22}$Na with HO frequencies for different $N_{max}$ are shown corresponding to two different $NN$ interactions.}
    \label{gs_entangle}
\end{figure}

% \begin{figure}[b]
% % \centering
%     \includegraphics[scale = 0.60]{22Na_entanglement1.pdf}
%     \caption{The variation of calculated g. s. $S_{tot}$ of $^{22}$Na with HO frequencies for different $N_{max}$ are shown corresponding to two different $NN$ interactions.}
%     \label{gs_entangle}
% \end{figure}

Additionally, we calculated the total entanglement entropies ($S_{tot}$) of $^{20}$Ne (0$^+$) and $^{22}$Na (3$^+_1$) for different NCSM parameters and the results are shown in Fig. \ref{gs_entangle}. For $^{20}$Ne, while considering $N_{max}$ = 0 space or up to $N$ = 2 major shells, the $S_{tot}$ for INOY (N$^3$LO) at $\hbar \Omega$ = 8 MeV is 2.739 (2.685). As the $\hbar \Omega$ increases, $S_{tot}$ increases slightly, reaching a maximum value of 3.838 (3.852) at $\hbar \Omega$ = 18 MeV and then decreasing. 
%4.385 (4.452). As the $\hbar \Omega$ increases, $S_{tot}$ increases slightly, reaching a maximum value of 4.471 (4.535) at $\hbar \Omega$ = 14 MeV and then decreasing. 
At $N_{max}$ = 0 model space, the $N$ = 0 and 1 orbitals are fully occupied, contributing 0 to the $S_{tot}$. As the model space increases to $N_{max}$ = 2, the $S_{tot}$ for INOY (N$^3$LO) become 5.387 (5.343) at $\hbar \Omega$ = 8 MeV. The differences between $S_{tot}$ ($N_{max}$ = 2) and $S_{tot}$ ($N_{max}$ = 0) are mainly contributed by $N$ = 0 and 1 orbitals along with $N >$ 2 orbitals. The contributions from $N$ = 2 orbitals also increase slightly as $N_{max}$ increases from 0 to 2. When the model space is increased further to $N_{max}$ = 4, the $S_{tot}$ becomes 6.584 (6.598) at the same HO frequency. More contributions from $N$ = 0 to 4 orbitals are seen towards $S_{tot}$ except for $N$ = 2 orbitals as compared to $N_{max}$ = 2 results. A similar behavior of the $S_{tot}$ is also seen for the odd-odd nucleus, $^{22}$Na, as shown in the second panel of Fig. \ref{gs_entangle}. In general, the increment of $N_{max}$ increases the $S_{tot}$, while the increment of $\hbar \Omega$ decreases $S_{tot}$ for a particular value of $N_{max}$ other than the $N_{max}$ = 0 model space. The $S_{tot}$ for $^{22}$Na is seen to be higher than $^{20}$Ne while considering same $\hbar \Omega$ and $N_{max}$. On the other hand, it is less dependent on the interactions employed for both the nuclei considered in this work. The $^{20}$Ne g.s. $S_{tot}$ attained a maximum at $\hbar \Omega$ = 10 MeV for both INOY and N$^3$LO interactions. Contrary to these results, the $^{22}$Na g.s. showed maximum at two different $\hbar \Omega$ values, 8 and 10 MeV, corresponding to INOY and N$^3$LO, respectively. We considered these frequencies as our second set of optimal frequencies. The NCSM calculations for low-energy spectra and a few electromagnetic transitions are performed at two different frequencies for each interaction: (i) at the optimal frequency based on minimum g.s. energies which are 20 MeV for INOY and 14 MeV for N$^3$LO interaction (set 1), (ii) at that frequency where g.s. $S_{tot}$ is the maximum for $N_{max}$ = 4 model space, as shown in the third subplot of Fig. \ref{gs_entangle} (set 2).

% \begin{figure*}[b]
%     \centering
% \includegraphics[scale = 0.9]{22Na_spectra.eps}
%     \caption{Caption}
%     \label{22Na_spectra}
% \end{figure*}

\subsection{Low-energy spectra and single orbital entanglement:}
\begin{figure*}
    \centering
\includegraphics[scale = 0.7]{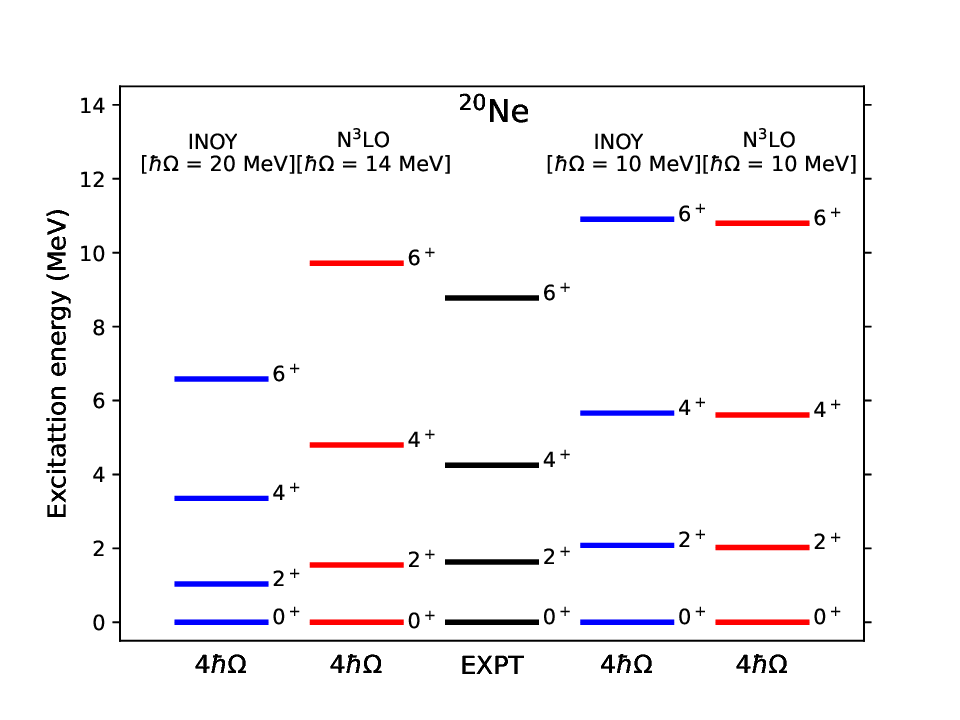}
\includegraphics[scale = 0.7]{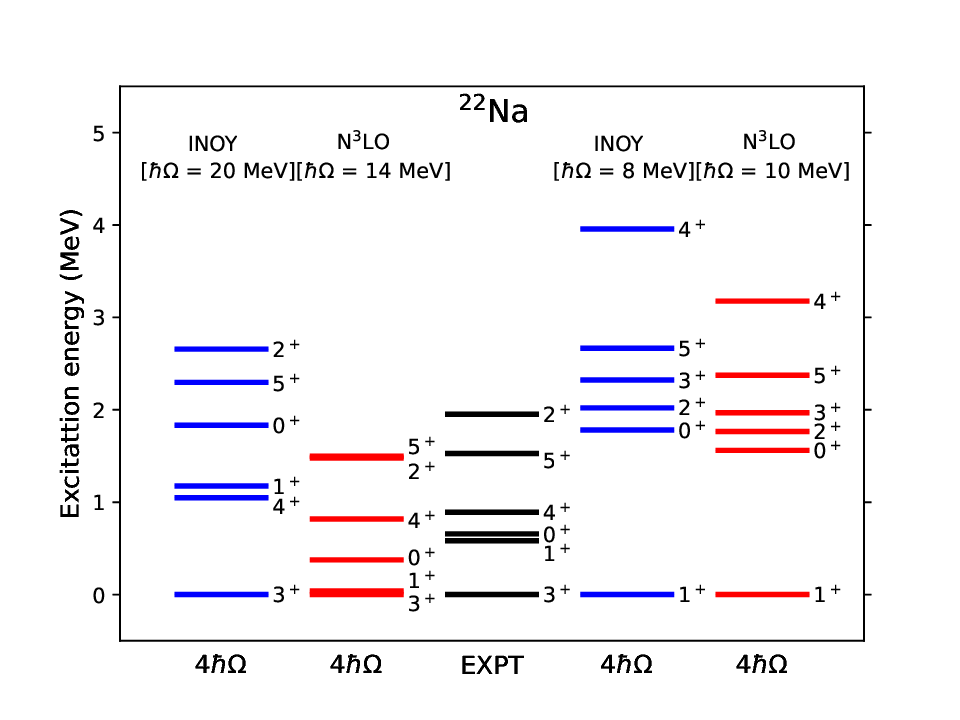}
    \caption{Low-energy spectra of $^{20}$Ne and $^{22}$Na are shown for two different interactions. All calculated results are corresponding to $N_{max}$ = 4.}
    \label{20Ne_spectra}
\end{figure*}

\begin{figure}[b]
    \centering
    \includegraphics[scale = 0.65]{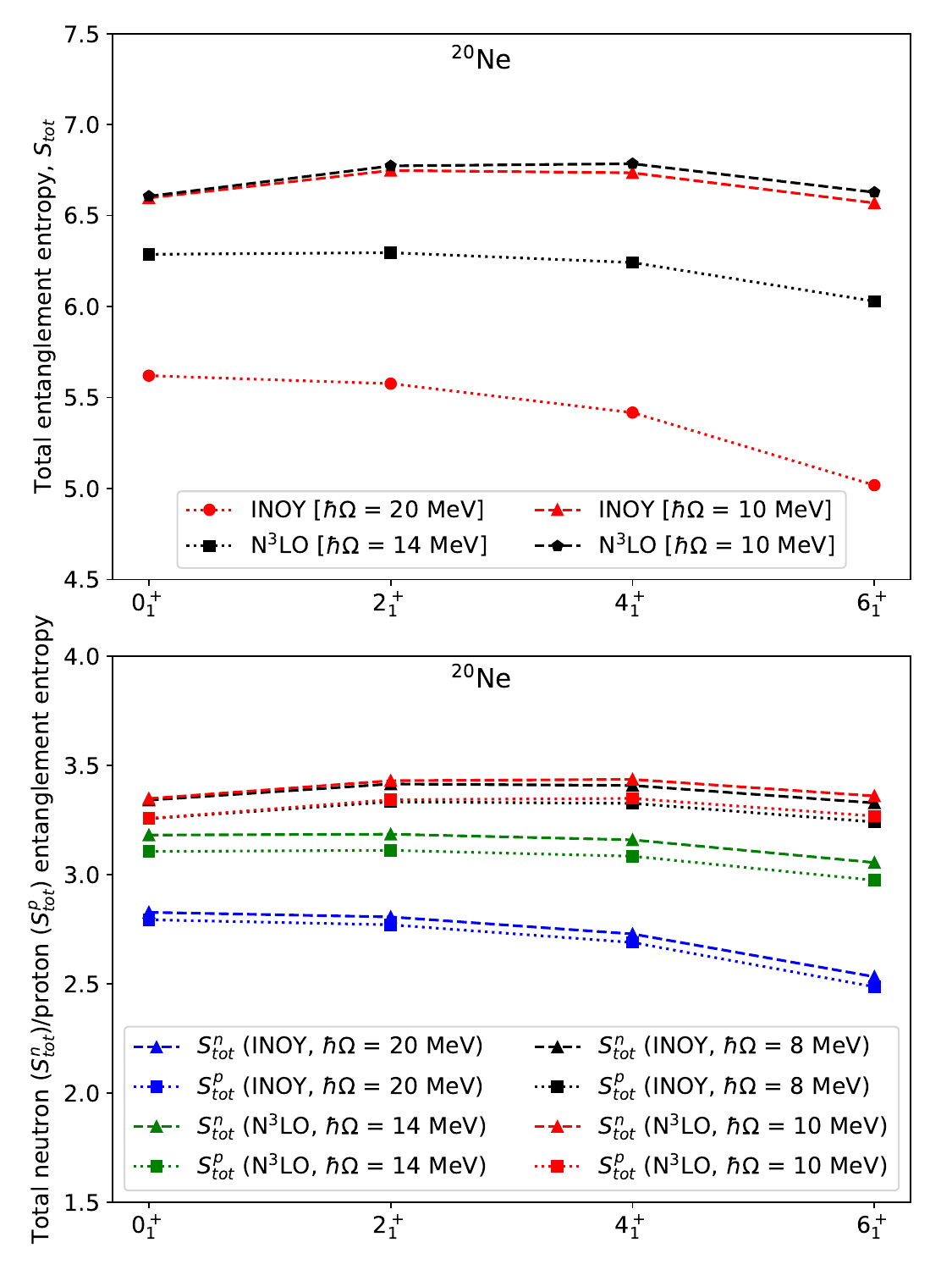}
    \caption{(a) Calculated $S_{tot}$ for four low-lying states of $^{20}$Ne using INOY and N$^3$LO interactions. The points connected by the dotted and dashed lines correspond to set 1 and set 2 optimal frequencies, respectively. (b) Calculated $S_{tot}^p$ (dotted line) and $S_{tot}^n$ (dashed line) for the four states of $^{20}$Ne are shown.}
    \label{20Ne_spectra_EN}
\end{figure} 

\begin{figure}
\hspace{2cm}\includegraphics[scale = 0.65]{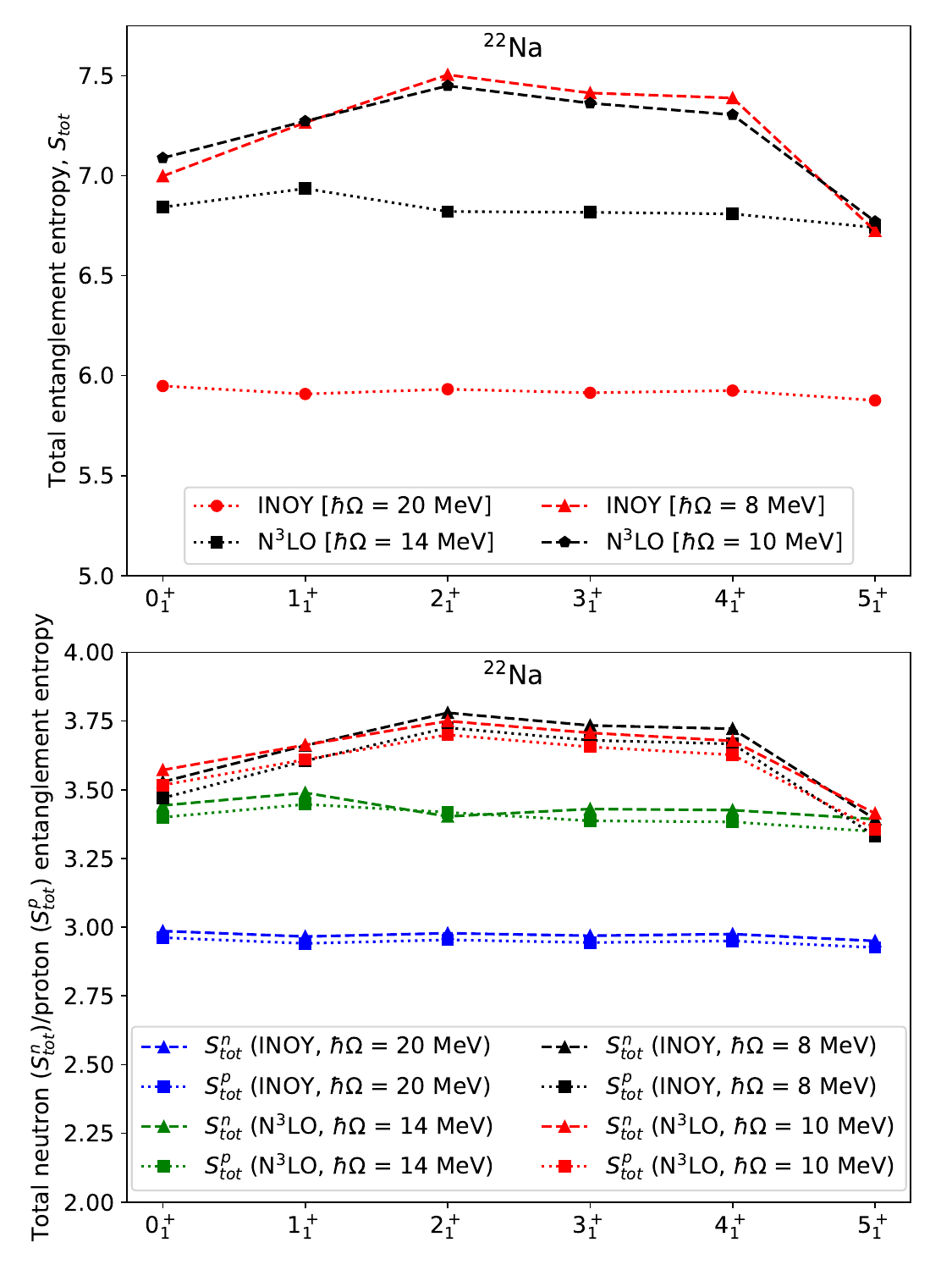}
     \caption{(a) Calculated $S_{tot}$ for six low-lying states of $^{22}$Na using INOY and N$^3$LO interactions. The points connected by the dotted and dashed lines correspond to set 1 and set 2 optimal frequencies, respectively. (b) Calculated $S_{tot}^p$ (dotted line) and $S_{tot}^n$ (dashed line) for the six states of $^{22}$Na are shown.}
     \label{22Na_spectra_EN}
 \end{figure}

We are now interested in the entanglement study in the low-energy spectra of $N = Z$ nuclei, $^{20}$Ne and $^{22}$Na. For the case of $^{20}$Ne, we are interested in four low-lying states: 0$^+_1$, 2$^+_1$, 4$^+_1$, and 6$^+_1$ and the calculated results are shown in the first panel of Fig. \ref{20Ne_spectra} along with the experimental data. The spectra results for set 1 are taken from Ref. \cite{chandan_Ne}, and for both sets, the ordering of these four states is well reproduced. However, the excited states corresponding to set 2 are higher than the experimental data as well as calculated data from set 1. Also, the set 2 spectra corresponding to INOY and N$^3$LO are nearly identical, unlike the case of spectra for set 1. In the case of $^{22}$Na, we are focusing on the lowest energy states having spin-parity between 0$^+$ and 5$^+$, and the results are shown in the second panel of Fig. \ref{20Ne_spectra}. From all the set 1 results presented in the figure, it can be seen that both interactions are able to reproduce correct g.s. of $^{22}$Na. Although the first excited state (1$^+_1$) for N$^3$LO interaction is found to be nearly degenerate with the g.s., it is able to reproduce the correct ordering up to 4$^+_1$ state. On the other hand, the set 2 calculated results show 1$^+_1$ as the g.s. instead of 3$^+_1$. 

Finally, we calculated the $S_{tot}$ for all ten states considered in this work using INOY and N$^3$LO interactions at $N_{max}$ = 4 model space. In the first panel of Fig. \ref{20Ne_spectra_EN}, the $S_{tot}$ for four low-lying states of $^{20}$Ne are shown corresponding to two different sets of optimal frequencies. The points connected by the dotted lines correspond to the $S_{tot}$ calculated at the set 1 optimal frequency, where the $S_{tot}$ for INOY interaction are, on average, 14.9 \% lesser than the N$^3$LO results. On the other hand, the points connected by the dashed lines represent the $S_{tot}$ calculated corresponding to the set 2 optimal frequency. Although the calculated $S_{tot}$ for the second set of optimal frequencies is higher than the first set, the $S_{tot}$ for INOY and N$^3$LO are close for the former set, suggesting that the calculated $S_{tot}$ is more dependent on $\hbar \Omega$ than the interactions employed. Additionally, in the second panel of Fig. \ref{20Ne_spectra_EN}, the total neutron (proton) single-orbital entanglement entropies, $S_{tot}^n$ ($S_{tot}^p$), which are the sum of neutron (proton) $S_i$ are shown. From the figure, it can be seen that $S_{tot}^p$ is slightly lower than the $S_{tot}^n$ in most cases. This could be due to the Coulomb and the charge symmetry breaking (CSB) parts of the interactions used, due to which $V_{pp}$ and $V_{nn}$ parts become different. In the first panel of Fig. \ref{22Na_spectra_EN}, the $S_{tot}$ for six low-lying states of $^{22}$Na are shown corresponding to two different sets of optimal frequencies. From the figure, it can be seen that the N$^3$LO interaction shows more total single-orbital entanglement than the INOY interaction, and it is, on average, 15.4 \% higher than the INOY results for the set 1 optimal frequency. The $S_{tot}$ for each state is almost the same while using INOY interaction contrary to the slightly varying $S_{tot}$ for N$^3$LO interaction peaking for the 1$^+_1$ with $S_{tot}$ = 6.936. On the other hand, no such deviations in $S_{tot}$ are seen between the INOY and N$^3$LO results while using the set 2 optimal frequency. However, for the later set of optimal frequency, low-energy states show deviations in $S_{tot}$, peaking at 2$^+_1$ and forming a minima at the 5$^+_1$ state. We also calculated the $S_{tot}^n$ ($S_{tot}^p$) for those states of $^{22}$Na, just like for the case of $^{20}$Ne and the results are shown in the second panel of Fig. \ref{22Na_spectra_EN}. From the figure, it can be seen that $S_{tot}^p$ is again slightly lower than the $S_{tot}^n$ in most cases, similar to the observation for the low-lying states of $^{20}$Ne. 

\subsection{Electromagnetic observables:} 
In this section, we are discussing the $E2$ and $M1$ transition strengths for a few transitions within the low-energy spectra of $^{20}$Ne and $^{22}$Na. For the case of $^{20}$Ne, we are mainly focussing on two $E2$ transitions: $B(E2; 2^+_1 \to 0^+_1$) and $B(E2; 4^+_1 \to 2^+_1$). On the other hand, for $^{22}$Na, we are interested in two $E2$ transitions: $B(E2; 1^+_1 \to 3^+_1$) and $B(E2; 5^+_1 \to 3^+_1$) and one $M1$ transition: $B(M1; 0^+_1 \to 1^+_1$). The NCSM results using INOY and N$^3$LO interactions for these transitions are shown in Table \ref{em} along with the experimental data from \cite{NNDC}. The NCSM calculations are performed at two different sets of optimal frequencies for each interaction. Firstly, we discuss the results corresponding to the first set optimal frequencies of each interaction as shown in the third and fourth columns of Table \ref{em}. The electromagnetic transition strengths, as shown in the table for $^{20}$Ne and $^{22}$Na corresponding to this set of optimal parameters, are taken from Refs. \cite{chandan_Ne} and \cite{chandan_Na}, respectively. 
	\begin{table}[t]
		%\centering
		\caption{\label{em} The electromagnetic transition strengths are calculated corresponding to the highest $N_{\mbox{max}}$. The $E2$ and $M1$ transitions are in  $e^{2}$ fm$^{4}$ and $\mu_{N}^{2}$ respectively. Experimental values are taken from Refs. \cite{NNDC}.  }
  %\begin{center}
		\begin{tabular}{|c|c|c|c|c|c|}
			\hline
   			$^{20}$Ne & EXPT & INOY & N$^3$LO & INOY  & N$^3$LO  \\
                 &           & [20 MeV] &  [14 MeV]& [10 MeV] & [10 MeV] \\
			%\hline
			\hline 
			$B(E2;2_{1}^{+}$ $\rightarrow$ $0_{1}^{+}$) & 65.4(32) &  10.35 & $<$ 0.01 & 35.19 & 35.67\\
			$B(E2;4_{1}^{+}$ $\rightarrow$ $2_{1}^{+}$) & 71.0(64) & 12.57 & $<$ 0.01 & 44.41 & 45.48 \\
			\hline
			$^{22}$Na & EXPT & INOY & N$^3$LO & INOY  & N$^3$LO  \\
                 &           & [20 MeV] &  [14 MeV]& [8 MeV] & [10 MeV] \\
			%\hline
			\hline 
			$B(E2;1_{1}^{+}$ $\rightarrow$ $3_{1}^{+}$) & 0.03457(29) &  $< $ 0.01 & 0.60 & 118.14 & 78.74\\
			$B(E2;5_{1}^{+}$ $\rightarrow$ $3_{1}^{+}$) & 19.0(15) & 3.87 & 7.15 & 0.81 & 2.75 \\
			$B(M1;0_{1}^{+}$ $\rightarrow$ $1_{1}^{+}$) & 4.96(18) & 4.37 &  12.60 & 16.32 & 15.97\\
			\hline
		\end{tabular}
  %\end{center}
	\end{table}

It is well known that the optimal frequencies for different nuclear observables can be different than the optimal frequencies evaluated based on the g.s. binding energy. The $S_{tot}$ is a quantity that is directly dependent on the nuclear wavefunctions, and based on the variation of $S_{tot}$ with NCSM parameters, another set of optimal frequencies is decided as discussed in Sect. \ref{sect3}. Now, we discuss the electromagnetic transition strengths calculated corresponding to the set 2 optimal frequencies. The new set of optimal frequencies is 10 MeV for both INOY and N$^3$LO in the case of $^{20}$Ne, and they are 8 MeV and 10 MeV, respectively, while considering the $^{22}$Na nucleus.  The electromagnetic transition strengths calculated corresponding to these frequencies are shown in the fifth and sixth columns of Table \ref{em}. The $E2$ transition strengths for $^{20}$Ne show significant improvement for this new set of optimal frequencies compared to the set 1 results; however, those results are still away from the experimental data. On the other hand, the $B(E2; 1^+_1 \to 3^+_1$) values calculated for the new set of optimal frequencies are significantly higher than the experimental data as well as the earlier calculated results. Also, the $B(M1; 0^+_1 \to 1^+_1$) corresponding to the new set of optimal frequencies increases considerably. However, the $B(E2; 5^+_1 \to 3^+_1$) decreases for this set of calculations. In order to establish a relationship between the $S_{tot}$ and the electromagnetic transitions among the low-lying states, we plot $S_{tot}$ for the low-lying states of $^{20}$Ne and $^{22}$Na in the first panels of Figs. \ref{20Ne_spectra_EN} and \ref{22Na_spectra_EN} corresponding to both sets of optimal frequencies. From the figures, it can be seen that the $S_{tot}$ for the new set of optimal frequencies is significantly larger than the initial ones. Also, it is noticed that the $S_{tot}$ between the initial and final states differ slightly where $B(E2; 2^+_1 \to 0^+_1$) and $B(E2; 4^+_1 \to 2^+_1$) for $^{20}$Ne and $B(E2; 1^+_1 \to 3^+_1$) and $B(M1; 0^+_1 \to 1^+_1$) for $^{22}$Na show significant increment. On the other hand, a significant difference in the $S_{tot}$ between the initial and final states in the case of $^{22}$Na nucleus is seen for $B(E2; 5^+_1 \to 3^+_1$) where a reduction in the $E2$ transition strength is observed. So, it can be concluded that large and comparable $S_{tot}$ in the initial and final states corresponding to an electromagnetic transition can enhance the transition strength significantly.

\section{Conclusions:}
\label{sect4}
In this work, we have investigated the entanglement structure of $N = Z$ nuclei, $^{20}$Ne and $^{22}$Na based on single-orbital entanglement within the NCSM formalism corresponding to two realistic interactions, namely, INOY and N$^3$LO. Firstly, we showed the variation of g.s. binding energy with the NCSM parameters: $N_{max}$ and $\hbar \Omega$ and decided the optimal frequencies for each interaction (set 1). Secondly, we evaluated the variation of total single-orbital entanglement, $S_{tot}$ for the g.s. of 
$^{20}$Ne and $^{22}$Na with NCSM parameters. It is seen that while the $S_{tot}$ increases with the increase in $N_{max}$, it decreases with $\hbar \Omega$ after attaining a maximum. The optimal frequencies for nuclear observables can differ from those based on ground-state binding energy, with $S_{tot}$ guiding alternative frequency choices (set 2). We then calculated the low-energy spectra of $^{20}$Ne consisting of four low-lying states: 0$^+_1$, 2$^+_1$, 4$^+_1$, and 6$^+_1$ for both sets of optimal frequencies.
Similarly, we calculated low-energy spectra of $^{22}$Na, having spin-parity between 0$^+$ to 5$^+$. Additionally, we calculated the $S_{tot}$ for each state considered in this work, and it is seen that $S_{tot}$ for each state of $^{20}$Ne is higher for the second set of optimal frequencies than the first set. A similar observation can also be found for the odd-odd nucleus considered in this work. However, for both nuclei, the $S_{tot}$ for INOY and N$^3$LO are close while considering the second set of optimal frequencies, suggesting that the calculated $S_{tot}$ are more dependent on $\hbar \Omega$ than the interactions employed. Finally, we calculated a few $E2$ transitions and one $M1$ transition within the low-energy spectra of $^{20}$Ne and $^{22}$Na, comparing NCSM results using INOY and N$^3$LO interactions with experimental data. For the $^{22}$Na nucleus, the N$^3$LO interaction gives better results for $B(E2; 5^+_1 \to 3^+_1$), and INOY performs well for the $M1$ transition while considering the first set of optimal frequencies. Electromagnetic transition strengths calculated at the second set of optimal frequencies show significant increases in $B(E2; 1^+_1 \to 3^+_1$) and $B(M1; 0^+_1 \to 1^+_1$), but a decrease in $B(E2; 5^+_1 \to 3^+_1$). It can be concluded that when the initial and final states of an electromagnetic transition have large and comparable $S_{tot}$ values, the transition strength can be significantly enhanced.

% \vspace{0.2cm}
%%%%%%%%%%%%%%%%%%%%%%%%%%%%%%%%%%%%%%%%%%%%%%%%%%%%
%\section*{ACKNOWLEDGMENTS}
\section*{Acknowledgments}

 We acknowledge financial support from SERB (India), CRG/2019/000556. 
 We would like to thank  Petr Navr\'atil for providing us with his $NN$ effective interaction code and Christian Forss\'en for making available the pAntoine \cite{pAntoine2,pAntoine3} code.

\end{document}